\documentclass[a4paper,fleqn,usenatbib]{mnras}

\usepackage[T1]{fontenc}
\usepackage{ae,aecompl}

\usepackage{graphicx}	
\usepackage{amsmath}	
\usepackage{amssymb}	

\title[Search For Type Ia Supernova NUV-Optical Subclasses]
{Search For Type Ia Supernova NUV-Optical Subclasses}

\author[David Cinabro et al.]{
David Cinabro$^{1}$\thanks{E-mail: david.cinabro@wayne.edu},
Daniel Scolnic$^{2,}$$^{3}$, Richard Kessler$^{2,}$$^{4}$, Ashley Li$^{2}$, and
Jake Miller$^{1}$ 
\\
$^{1}$Wayne State University, Department of Physics and Astronomy, Deroit, MI, USA\\
$^{2}$University of Chicago, Kavli Institute for Cosmological Physics, Chicago, IL, USA\\
$^{3}$Hubble, KICP Fellow\\
$^{4}$Department of Astronomy and Astrophysics, University of
Chicago, Chicago, IL, USA
}

\date{Accepted XXX. Received XXX; in original form XXX}

\pubyear{2016}

\begin{document}
\label{firstpage}
\pagerange{\pageref{firstpage}--\pageref{lastpage}}
\maketitle

\begin{abstract}
In response to a recently reported observation of evidence for two classes of 
Type Ia Supernovae (SNe~Ia) distinguished by their brightness in the rest-frame 
near ultraviolet (NUV), we search for the phenomenon in publicly available 
light-curve data.
We use the SNANA supernova analysis package to simulate
SN~Ia-light curves in the Sloan Digital Sky Survey Supernova Search (SDSS)
and the Supernova Legacy Survey (SNLS) with a model of two
distinct ultraviolet classes
of SNe~Ia and a conventional model with a
single broad distribution of SN-Ia ultraviolet brightnesses.
We compare simulated distributions of rest-frame colors with
these two models to those
observed in 158 SNe~Ia in the SDSS and SNLS data.
The SNLS sample of 99 SNe~Ia is in
clearly better agreement with a model with one class of SN~Ia light curves and shows no
evidence for distinct NUV sub-classes.  The SDSS sample of 59
SNe~Ia with poorer color resolution does not distinguish 
between the two models.
\end{abstract}



\section{Introduction}
\label{sec:intro}

A recent claim in~\citet{Milne2015} is that
there are two distinct groups of Type Ia Supernovae (SNe~Ia)
distinguished by their brightness in rest frame near ultraviolet.
The light curves of the two sets, red and blue, differ 
by $0.46$~mag in their $u-v$ colors at the time of
peak brightness in the rest-frame B band ($B$-peak) where the color is derived from 
spectrophotemetry of 23 SNe~Ia observed with
the UVOT instrument on the SWIFT satellite~\citep{SWIFT} and
52 SNe~Ia observed with Keck and the VLT with
spectrophotometry matched to the UVOT system.  The two sets
have a $0.34$~mag difference in the $u-b$ color and an
insignificant difference in the $b-v$ color at peak brightness.  Further, the 
fractions of SN~Ia in the
red and blue sets change as a function of redshift with the red
sample dominant (60-70\% of the total) at redshifts smaller than 0.1,
and the blue sample dominant (80-90\% of the total) at redshifts from 0.4 to 1.0.  
The study also found SNe~Ia in the red sample 
exhibited larger ejecta velocities in their spectral features.
\citet{Milne2015} claims
that SNe~Ia light curve fits for the red sample
will underestimate the host galaxy extinction, leading
to a redshift dependent bias in the corrected peak brightness
and also in the inferred cosmological parameters.

The importance of SNe~Ia in modern cosmology cannot be understated.
Observations just before the turn of the century provided the first
clear evidence of an accelerated expansion of the 
universe~\citep{Reiss, Perlmutter} and subsequent observations combined
with the clustering of galaxies and the cosmic microwave background
point towards this acceleration being caused by a negative pressure
fluid, dubbed Dark Energy~\citep{PDGdarkenergy}.  These observations
and the precision of the extraction of cosmological parameters inferred
from SNe~Ia
depend on the assumption that the light curves of SNe~Ia at low redshift
are standardized in the same way as SNe~Ia at high redshift.  The
observations of \citet{Milne2015} call this underpinning assumption 
into question. 

The modeling of the rest frame ultraviolet 
brightnesses of SNe~Ia is more difficult than in the visible.
SNe~Ia are dimmer in the rest frame ultraviolet than
the visible, the effects of extinction are larger in the
ultraviolet, and ground based observations are difficult
with large and highly variable ultraviolet atmospheric absorption.
The Joint Lightcurve Analysis~\citep{jla} (JLA) is the largest
sample to date used for SN~Ia cosmology, and they have publicly released
many high quality SN~Ia light curves.  In the JLA, the ultraviolet 
behavior of SN~Ia is empirically modeled with a distribution of brightnesses that is much broader
than the distribution in the visible.
This model does not agree with the \citet{Milne2015}
observation of two narrow and distinct distributions in the ultraviolet.

Checking the claim of \citet{Milne2015} with the JLA data is not as
simple as it might seem.  At low redshift, the JLA sample is
made of a heterogeneous collection of SN~Ia light curves
observed by many different instruments and surveys.
Unknown selection effects for this sub-sample make it difficult 
to interpret \citep{Scolnic2014}.
To avoid possible biases from selection effects, we use
the SDSS-II \citep{Frieman2008} and SNLS \citep{Astier06} 
SN~Ia samples, whose selection effects are well modeled 
with Monte Carlo simulations \citep{KesslerInt,jla}.
However, these samples are not well suited to ultraviolet
photometry.  The SDSS $u-$filter is less efficient than
the $g-$, $r-$, $i-$filters, and the survey found only
12 securely identified SNe~Ia at redshift smaller below 0.1
which could be used to directly check the \citet{Milne2015} observation.
The SNLS data does not
provide ultraviolet filter photometry.  With insufficient
ultraviolet SN~Ia data in the public domain, a more promising approach is to look
at higher redshift where the rest frame ultraviolet is observed
in the visible.  From redshift of 0.3 to 0.7, the spectral range of
the rest frame UVOT $u-$, $v-$, and $b-$filters is observed in the 
SDSS and SNLS $g-$, $r-$, and $i-$filters.

The next section describes the models of SN~Ia light curves that
we used to check the observations in \citet{Milne2015}.  
Section~\ref{sec:simulation} describes
our simulations of the SDSS-II and SNLS surveys, and the 
predictions based on our model of the \citet{Milne2015} observation.
In Section~\ref{sec:compare} we
compare the observed SN~Ia light curves with the two models, and we end with a short
conclusion.

\section{SN~Ia Models}
\label{sec:models}

We use the publicly available SNANA~\citep{SNANA}\footnote{http://snana.uchicago.edu}
package to perform the analysis
described below.  SNANA is a supernova light curve analysis package
that allows for detailed simulations of SNIa light curves 
for arbitrary instruments, cadences and observing conditions.
To develop a model for the \citet{Milne2015} observations, 
we simulate SNIa light curves with infinite
photon statistics observed with an error free
instrument at redshift of 0.01, called ``perfect mode'',
to view and compare rest-frame models without 
instrumental effects.
We observe the time dependence of the colors and $B$-peak colors in the
filters on the UVOT instrument~\citep{UVOT}.  Our goal
here is to develop a SN~Ia light curve model that reproduces
the Milne observations and compare it to an existing, more conventional
description of SNe~Ia ultraviolet
brightness.  

We simulate light curves in the UVOT passbands to 
compare with \citet{Milne2015}, and also in the SDSS and SNLS
passbands, as described in the JLA~\citep{jla}, to compare with data.
The filters of the SDSS and SNLS are similar, but not exactly the same.  Table~\ref{tab:filters}
gives the central wavelengths of the filters we reference in this work.
\begin{table}
\caption{Central wavelengths of the filters used in this work.  Details of the UVOT filters can be found in~\citet{UVOT} and for the SDSS and SNLS filters in~\citet{jla}.}  The UVOT filters are similar to the standard Bessel filters.
\begin{tabular}{l|l}
Name          & Central Wavelength (\AA) \\ \hline
UVOT-$u$      & 3465 \\ 
UVOT-$b$      & 4392 \\ 
UVOT-$v$      & 5468 \\
SDSS/SNLS-$g$ & 4760 \\
SDSS/SNLS-$r$ & 6230 \\
SDSS/SNLS-$i$ & 7630 \\
SDSS/SNLS-$z$ & 9130
\end{tabular}
\label{tab:filters}
\end{table}

\begin{figure}
	\includegraphics[width=\columnwidth]{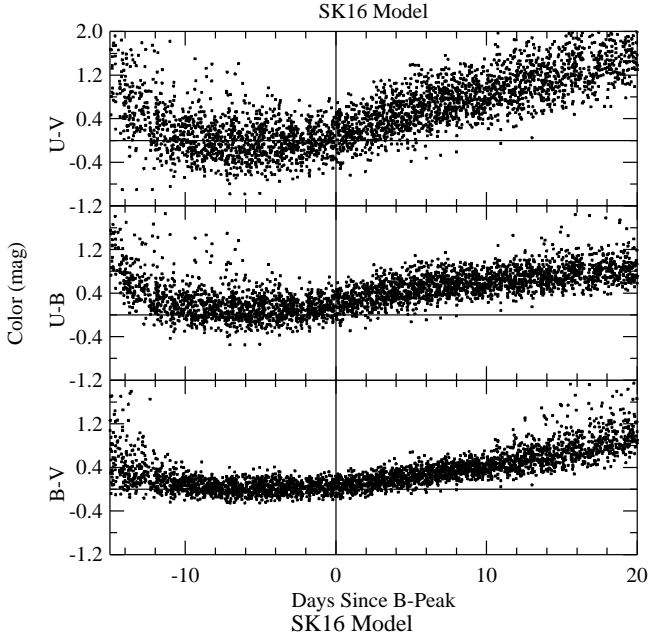}\\
	\includegraphics[width=\columnwidth]{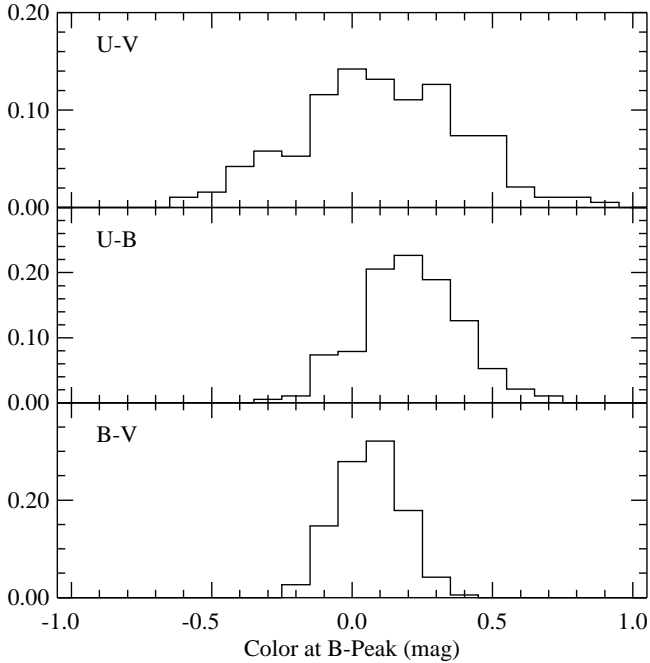}
    \caption{Rest frame colors in the UVOT filters for the SK16 
model in SNANA perfect mode as described in the text of SN~Ia
versus the time since the $B$-peak (top)
and within one day of the $B$-peak (bottom).}
    \label{fig:saltiilc}
\end{figure}

Our base model for SN~Ia light curves is the SALT-II model~\citep{salt}.  
The model describes the shape of light curves with two 
parameters, $x_1$, which gives the width-luminosity relation,
and $c$, the color parameter which parametrizes intrinsic SN~Ia colors
and the effect of extinction.  
The parent population of these parameters used
in our conventional simulation follow those measured by Scolnic and Kessler~\citep{ska16},
and thus we call this the SK16 model.   
Figure~\ref{fig:saltiilc} shows
the time dependence of the rest frame colors in the UVOT filters and within 
one day of $B$-peak in this SK16 model.
The model includes the
effect of Milky Way extinction from \citet{SFD98}.  
The observed scatter is due to a combination of intrinsic 
brightness variations, the underlying population of color 
($c$) and stretch ($x_1$), and Milky Way extinction.
In this model the ultraviolet
part of the spectrum shows a larger scatter
than in the visible.

We develop a model of the \citet{Milne2015} observation by 
introducing two classes of light curves distinguished
by their brightnesses in the ultraviolet part of the spectrum.
The resolution of the UVOT photometry, 
the red and blue histograms in Figure~2 of \citet{Milne2015}
showing the $u-v$-color for example, is between 0.05 and 
0.07 mag~\citep{MilnePrivate}.  Given the observed 
width of the two color peaks, in the range
of 0.07-0.09 mag, this implies the contribution to the width due to
the properties of the observed SN~Ia (the stretch distribution and 
Milky Way extinction) is very narrow, less than 0.05 mag.

To reproduce the observed 
features we modify the base prediction of the SALT-II
rest frame spectral flux, $F$, in the wavelength range 2700 \AA ~to 4300 \AA ~by
\begin{equation}
F = F^{(-0.4 \mathrm{dm})},
\label{eq:bifurcation1}
\end{equation}
where
\begin{equation}
\mathrm{dm} = \pm \left(\frac{0.55}{2}\right) \sin{\left(\pi \frac{(\lambda - 2700 \mathring{\mathrm{A}})}{1600 \mathring{\mathrm{A}}}\right)} (1 + 0.04 {\mathrm{R_G}) \mathrm{mag}.
\label{eq:bifurcation2}
}\end{equation}
The wavelength, $\lambda$, is in Angstroms, and $\mathrm{R_G}$ is a Gaussian distributed random number with standard deviation of one.  The 
amplitude and wavelength range of this bifurcation is
chosen to match the color separations in \citet{Milne2015}.
The amplitude is larger than the observed $u-v$ color separation
as it represents the maximum separation at only one wavelength,
while the color comes from integrating over the wavelength 
range of the $u-$filter.  The size of the Gaussian smearing
reproduces the narrow width of the NUV peaks.  Over the 
entire wavelength range there is an additional
coherent scatter drawn from a Gaussian with a width 0.08~mag
matching the width of the SNe~Ia brightness distribution.  
Those light curves with
the brighter ultraviolet distortion are 
the ``blue'' sample and those with the dimmer are the ``red'' sample,
and a random selection between the two is made for each light curve.

\begin{figure}
	\includegraphics[width=\columnwidth]{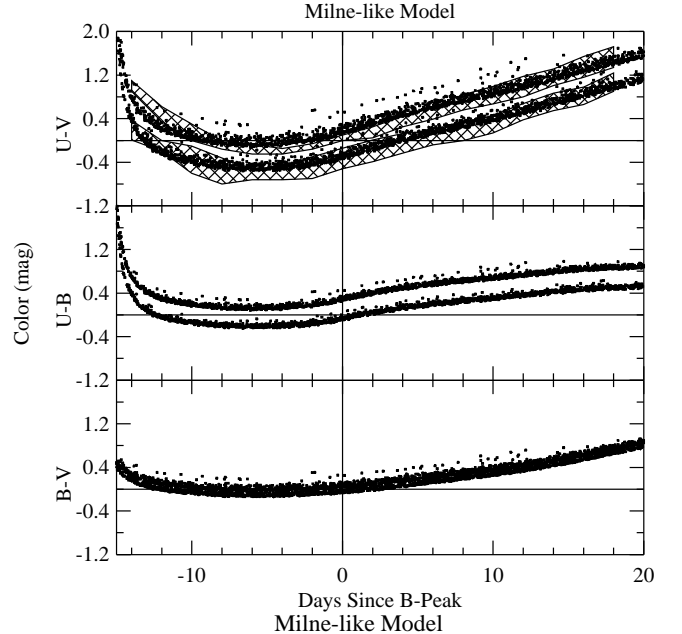}\\
	\includegraphics[width=\columnwidth]{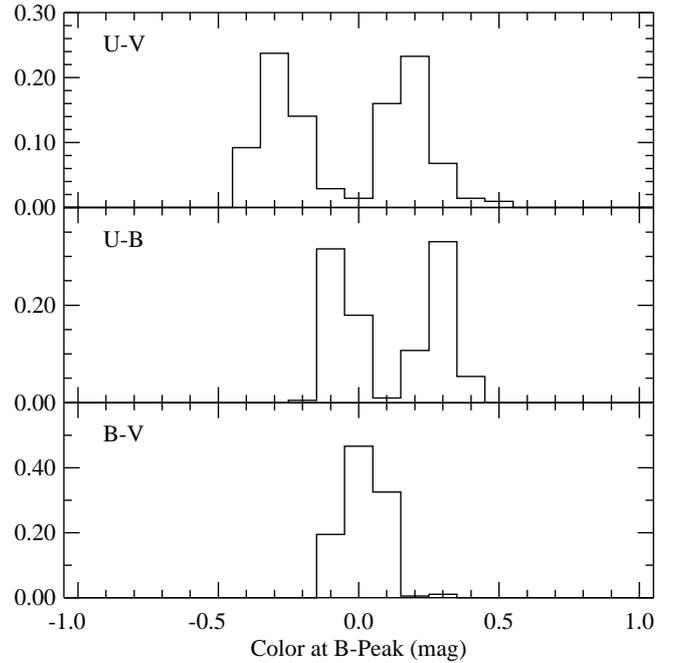}
    \caption{Using the SNANA ``perfect mode'' simulation for the Milne-like 
model, the top panel shows UVOT rest-frame colors versus time, 
and the bottom panel shows rest-frame colors within one day of $B$-peak.  In
the topmost panel showing the $u-v$-color versus time, the hatched regions
are the range of red, upper, and blue, lower, sample colors taken from 
Figure~1 of \citet{Milne2015}.}
    \label{fig:milnelc}
\end{figure}

We find that Equations~\ref{eq:bifurcation1} and~\ref{eq:bifurcation2} combined with 
non-zero values for the SALT-II $c$ parameter results in $u-v$ color peaks 
that are too broad to be consistent with
the \citet{Milne2015} observation.
To preserve the sharp color features, we set
the SALT-II color parameter, $c$, to zero.  With $c = 0$,
Figure~\ref{fig:milnelc}
shows time dependence of the rest frame colors in the UVOT filters 
and the colors within one day of $B$-peak in
this ``Milne-like'' model.
At $B$-peak, the width and separation of the red and 
blue peaks in the $u-v$ and $u-b$ colors
agrees well with \citet{Milne2015}.
The population of the two classes is equal in these illustrations.
This model reproduces well the data displayed in Figures 1-3, 9-12, and Table~2 
in \citet{Milne2015}.  It also incorporates the same variations of the SALT-II $x_1$
parameter and expected Milky Way extinction as we use in the SK16 model described
above.   

In the Milne-like model, variation in an extracted value of
the SALT-II color parameter away from zero is caused by the two different classes of 
SNe~Ia in the ultraviolet, red and blue, rather than an 
underlying intrinsic color population as described in \citet{ska16}.
Other choices for the parameters of the bifurcation given in \ref{eq:bifurcation2}
such as narrowing the wavelength range to 2700-3700~\AA~or
reducing the magnitude of the separation are discussed further
in Section~\ref{sec:compare}.

\section{Simulation of SN~Ia Surveys}
\label{sec:simulation}

Using the SK16 and Milne-like models, we simulate
SN~Ia light curves corresponding to the SDSS and SNLS
data in the JLA.  Our goal here is to use the JLA data
to compare against the two models when all observational
effects are included, and develop a method of choosing 
between the two models.  The simulations of
the SDSS and SNLS supernova surveys include the exact cadence of observations, photometric
uncertainties, redshift distributions, and spectroscopic identification
efficiencies observed in the surveys.  

To ensure robust light curve fits, we apply 
selection requirements to both the data and 
simulation.  We require at least
five epochs of observation in at least three of the $g-$, $r-$, $i-$, and $z-$bands.  
Observations must have a flux measured with signal to noise better than 3.0
in $g$, $r$, $i$ and 1.0 in $z$.  Light curves passing these criteria
are fit with the SALT-II model.  We only consider epochs within
$-15$ and $+50$ days of the fitted $B$-peak in the rest frame.
The results of the fit are accepted if there is at least one epoch before
and three after the $B$-peak, and the fit has a $\chi^2$ probability of 
greater than $0.001$.
We simulate samples that are about 15 times larger than the
data samples.  We consider SN~Ia with redshifts in the
range 0.3 to 0.4 for the SDSS and 0.3 to 0.7 for the SNLS.

These selections mirror the requirements on the JLA sample except
that we require the presence of $z$-band photometry.  The simulation
predicts that 9\% of SNe~Ia light curves in the JLA sample in the indicated 
redshift ranges will not pass to our sample.

\begin{figure*}
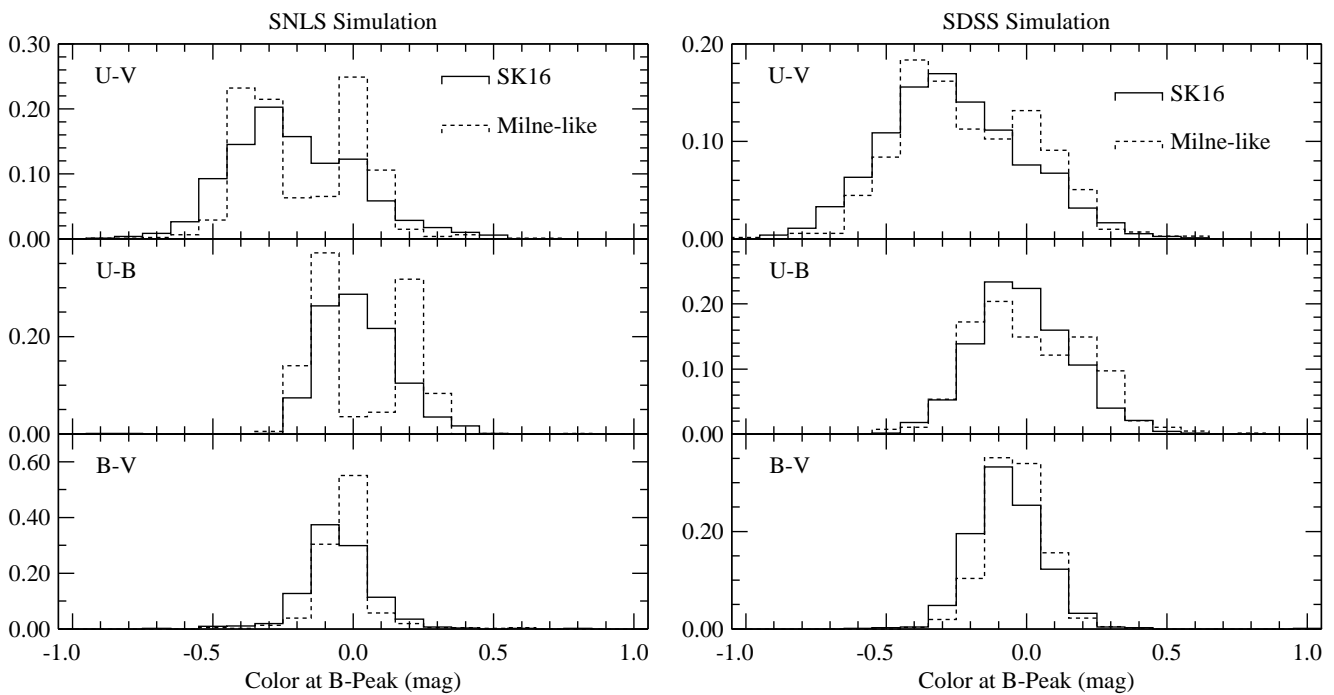

\begin{tabular}{cc}
\includegraphics[width=\columnwidth]{simcolors_a.epsi} &
\includegraphics[width=\columnwidth]{simcolors_b.epsi}
\end{tabular}
    \caption{The fitted values of the restframe colors at $B$-peak for the
SK16 and Milne-like models as described in the text for simulations
of the SNLS (left) and SDSS (right) surveys.}
    \label{fig:simcolors}
\end{figure*}

We extract the rest frame colors using the fitting procedure described in Section 4.3 of
\citet{KesslerInt}.  Briefly, after an initial fit to the SALT-II model, additional
constrained fits are done to the photometry results of the two observer frame filters
that most closely correspond to the best match in redshifted wavelength to the 
desired rest frame filter photometry.
The procedure introduces a negligible additional uncertainty on the
rest frame colors, and allows direct comparisons between the photometric
observations of SDSS and SNLS with the spectrophotometric observations in \citet{Milne2015}.
Figure~\ref{fig:simcolors} shows the distribution of colors at $B$-peak we
expect from the two surveys for the two models.  

The analysis of simulated SNe~Ia shows that we could clearly distinguish between the Milne-like model,
with two peaks clearly seen in the distribution of the $u-v$ color, and the SK16 model,
with the $u-v$ color distribution appearing as one broad peak, in the SNLS survey.  In the 
SDSS survey it is more difficult to see a difference with only a hint of two peaks
for the Milne-like model, but the SK16 model produces a slightly narrower distribution of the
$u-v$ and $u-b$ colors than the SK16 model.  The resolution of the 
$B$-peak colors of the SNLS is better than the SDSS.

\begin{figure}
\includegraphics[width=\columnwidth]{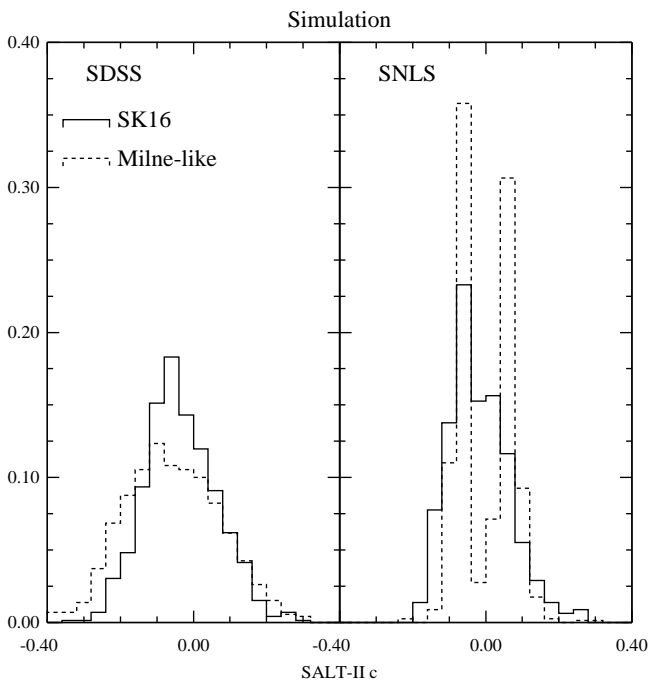}
    \caption{
For simulations of the SK16 (solid) and Milne-like (dashed) 
models, the fitted SALT-II color parameter ($c$) is shown
for SDSS (left) and SNLS (right).}
    \label{fig:simc}
\end{figure}

Further, we note that the fitted SALT-II color parameter is also
sensitive to the two models.  Figure~\ref{fig:simc} shows the distribution
of the fitted value for $c$ in the two simulated surveys for the two
models.  There is a clear difference between the two models.  The
SNLS simulation shows two narrow peaks for
the Milne-like model and one broad peak for the SK16 model.
The simulation of the SDSS shows a broader distribution for the
Milne-like model than the SK16 model.
The distribution of the fitted value of $c$ is sensitive to the 
different predictions of the two models.

We note that the rest frame colors are highly correlated with the 
fitted value of $c$ and considering only one of the distributions,
$c$ or $u-v$ color for example, is sufficient to discriminate
between the SK16 and Milne-like models for the brightness of
SNe~Ia in the rest frame ultraviolet.  

\section{Comparison of Survey Observations and Simulations}
\label{sec:compare}

We apply the analysis described above to the data from the SDSS and
SNLS supernova surveys in the JLA sample.  This only includes light
curves that have been
clearly typed with spectroscopy as SN~Ia,
and thus their redshifts are well measured.  Our analysis accepts
59 and 99 light curves in the SDSS and SNLS respectively.
After applying our redshift cuts to the JLA sample,
the other cuts reject 9 light curves ($5\pm2$\%), consistent
with the simulation prediction or 9\%.

\begin{figure*}
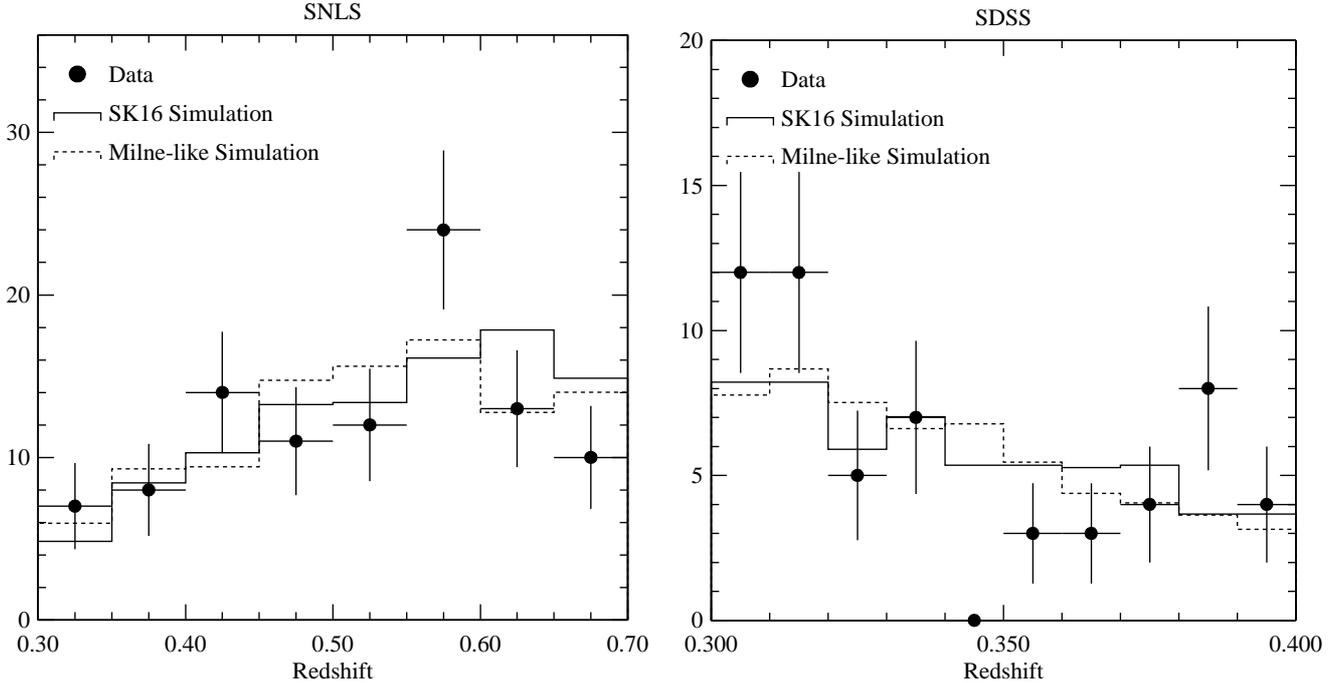

\begin{tabular}{cc}
\includegraphics[width=\columnwidth]{redshift_a.epsi} &
\includegraphics[width=\columnwidth]{redshift_b.epsi}
\end{tabular}
    \caption{The Redshift distribution of the SN~Ia accepted in
the analysis comparing the data and simulations of the SK16 and 
Milne-like models as described in the text in
the SNLS (left) and SDSS (right) surveys.}
    \label{fig:redshift}
\end{figure*}

\begin{figure*}
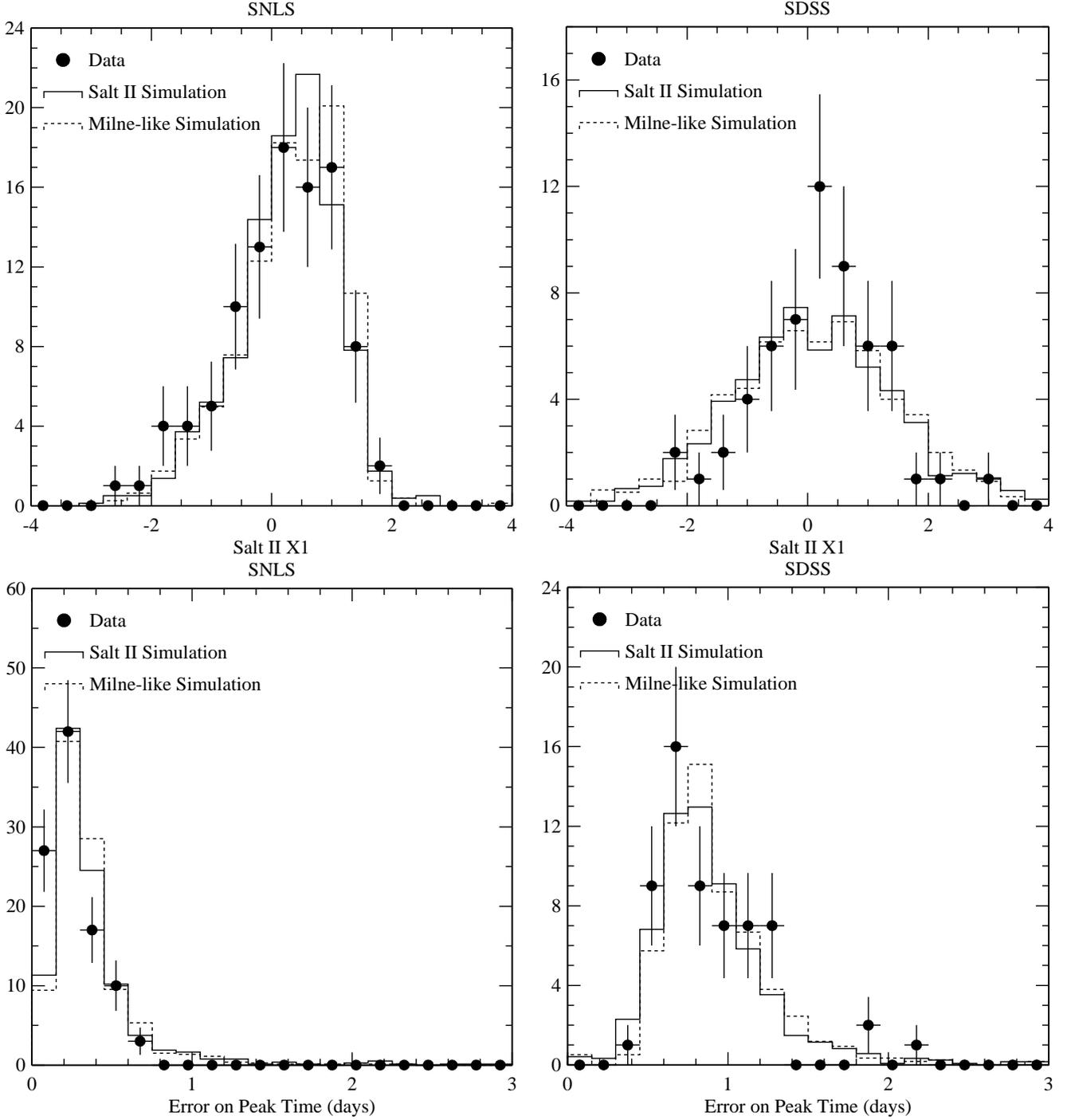

\begin{tabular}{cc}
\includegraphics[width=\columnwidth]{x1_a.epsi} &
\includegraphics[width=\columnwidth]{x1_b.epsi} \\
\includegraphics[width=\columnwidth]{deltat_a.epsi} &
\includegraphics[width=\columnwidth]{deltat_b.epsi} \\
\end{tabular}
    \caption{The distributions of the fitted values of
the SALT-II $x_1$, stretch, parameter (top) and
the uncertainty on the time of the $B$-peak (bottom)
comparing the data and simulations of the SK16 and 
Milne-like models as described in the text in
the SNLS (left) and SDSS (right) surveys.}
    \label{fig:x1anddeltat}
\end{figure*}

The redshift distributions
for the SN~Ia light curves accepted for analysis by both surveys agree well with
the simulation showing that efficiency and selection effects are well
modeled.  The distributions for the SALT-II $x_1$ parameter and the uncertainty
on the peak time also agree well between the simulation and the data
showing that the results of the SALT-II fits to the accepted light curves
are also well modeled by the simulation.  The uncertainty on the peak time
also contributes to the resolution of $B$-peak colors.
The simulated results for both the SK16 model and Milne-like model
in these parameters show no obvious dependence on
the two underlying SN~Ia light curve models.
Figures~\ref{fig:redshift}
and~\ref{fig:x1anddeltat} compare the results of our simulations
with the results of the data analysis for these parameters.

\begin{figure*}
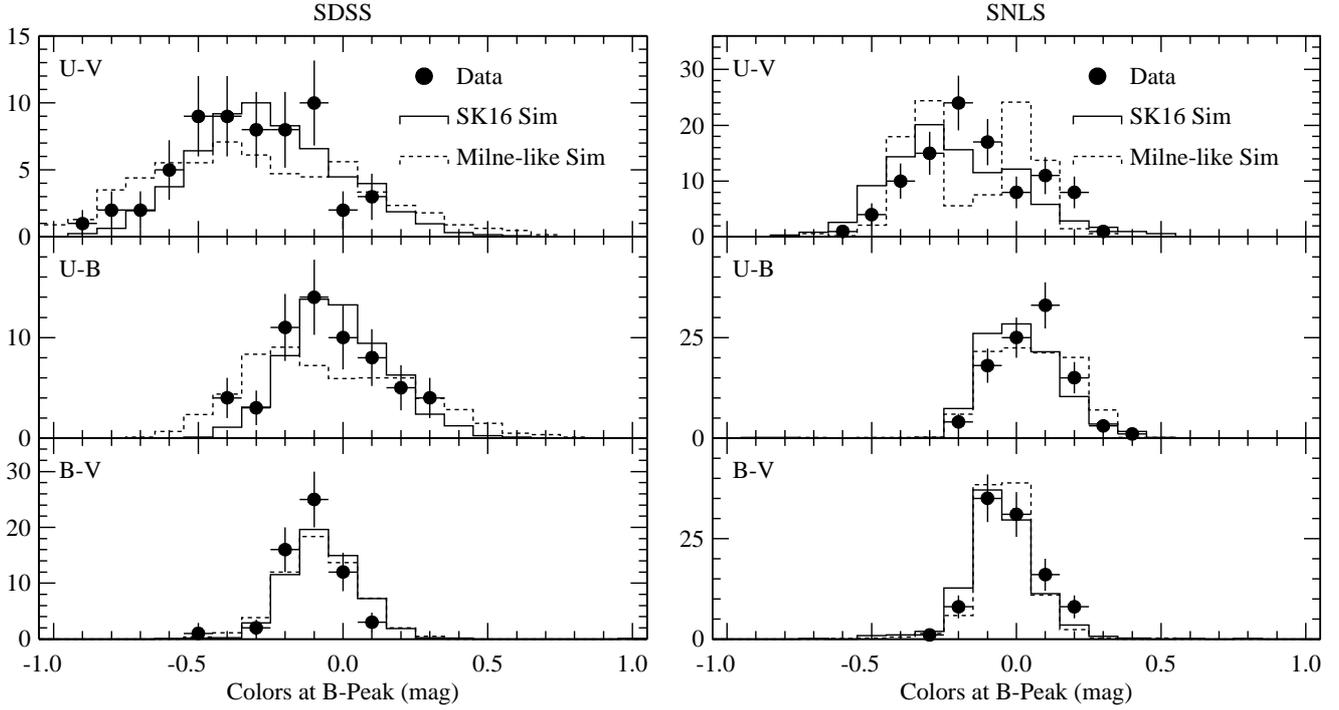

\begin{tabular}{cc}
\includegraphics[width=\columnwidth]{datacolors_a.epsi} &
\includegraphics[width=\columnwidth]{datacolors_b.epsi}
\end{tabular}
    \caption{The fitted rest frame $B$-peak colors
comparing the data and simulations of the SK16 and 
Milne-like models in
the SDSS (left) and SNLS (right) surveys.}
    \label{fig:datacolors}
\end{figure*}

Figure~\ref{fig:datacolors} compares the distribution of the 
fitted rest frame colors among the
SNLS and SDSS data with simulations of the SK16 and
Milne-like models.  Here the Milne-like model has 50\% red
and 50\% blue light curves.  
The colors at $B$-peak
clearly agree better with the SK16 model showing a
wide distribution in $u-v$ rather than two narrow peaks
in the SNLS as we would expect in the Milne-like model.
For the SDSS, the $u-v$ distribution is slightly
narrower compared to the Milne-like model.
The comparison is less
clear for the $u-b$ color where the Milne-like model distribution
is wider than the data, and the difference
between the SK16 and Milne-like models is smaller.  There
is no obvious difference between the data and the two models
in the $b-v$ color.

\begin{figure*}
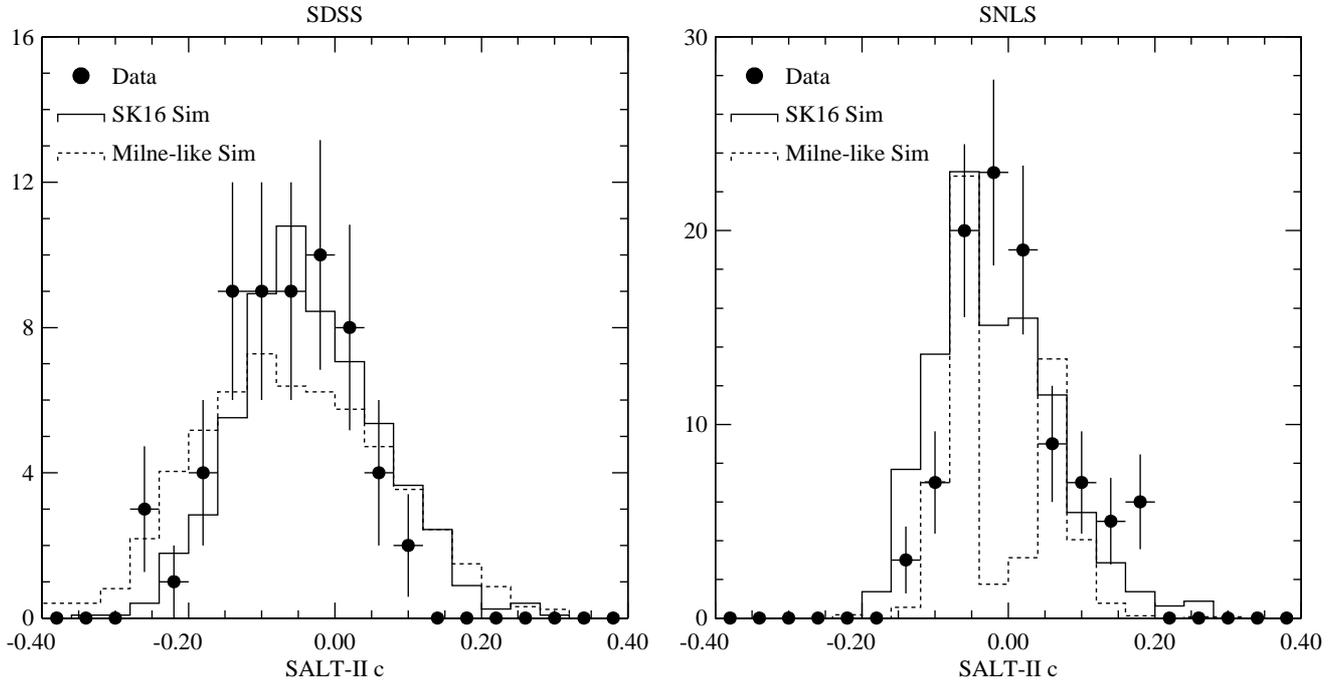

\begin{tabular}{cc}
\includegraphics[width=\columnwidth]{datac_a.epsi} &
\includegraphics[width=\columnwidth]{datac_b.epsi}
\end{tabular}
    \caption{The fitted SALT-II color parameter
comparing the data and simulations of the SK16 and 
Milne-like models in
the SDSS (left) and SNLS (right) surveys.}
    \label{fig:datac}
\end{figure*}

Figure~\ref{fig:datac} compares the distribution of the 
fitted SALT-II color parameter between the data and 
the simulations.  We considered three models
of this distribution to compare
with the data in a simple $\chi^2$ minimization.  
The first model is SK16, 
and it is fit to the data with fixed shape
allowing only the distribution's area to vary.  
The second is the Milne-like model
where the predictions for the fractions of blue and red
light curves
are taken from the observations displayed in Figure~4 of 
\citep{Milne2015}: roughly 70\% and 30\% respectively in the 
SDSS corresponding
to the redshift range of 0.3-0.4 and
80\% and 20\% in the SNLS in the redshift range 0.3-0.7.
We also fit this model to the data only allowing its area
to vary.  The third is a variant of the Milne-like model where
we allow the areas of the contributions from the red
and blue distributions to
vary independently.

In the SNLS data the SK16 model is decisively favored.
As can be seen on the right of Figure~\ref{fig:datac},
this model, the solid histogram, agrees reasonably
well with the data, dots, and gives $\chi^2 = 18$
for $8$ degrees of freedom.  Note that we did no
tuning of the parameters of this model to match our
data set, but simply used the parameters found in
a very different analysis done by Scolnic and 
Kessler~\citep{ska16}.  The Milne-like model
with fixed fractions of blue and red light curves fits
poorly giving $\chi^2 = 49$.  The Milne-like model
with floating fractions of blue and red light curves also
agrees poorly with the data giving $\chi^2 = 47$ for $7$ degrees of
freedom.  This fit has $(41\pm10)$\% of the light curves
from the blue sample, and is displayed as the dashed histogram
in the right panel of Figure~\ref{fig:datac}.  The relative
probability of the Milne-like model based on
the $\chi^2$ probability for these fits is 
smaller than $3 \times 10^{-6}$.

The results for the SDSS data, left side of Figure~\ref{fig:datac},
are not able to distinguish between the two models.
The fit to the SK16 model 
agrees well, giving a $\chi^2 = 7$ for $9$
degrees of freedom and is shown by the solid histogram.  
The fits to the Milne-like model with fixed
blue and red light curve fractions gives $\chi^2 = 12$,
and the Milne-like model with floating light curve
fractions gives $\chi^2 = 12$ for $8$ degrees of freedom
with $(49\pm16)$\% blue light curves.  This latter fit
is shown by the dashed histogram.  That the SDSS data
has poorer discrimination power than the SNLS data is not
surprising given the smaller number of SN~Ia in the SDSS sample
and poorer resolution on the peak colors than the SNLS.
Nevertheless the SDSS data show excellent agreement with
the SK16 model.

We explore variations of the Milne-like model consistent 
with the \citet{Milne2015} observation including
details of the bifurcation in the ultraviolet and the size of
the color separation.  We reduced the wavelength range
of the bifurcation between the red and blue
sample from 2700--4300~\AA~to 2700--3700~\AA~and varied the $u-v$
red and blue color separation in the range of 0.26-0.55 mag from the
central 0.46 mag.  Those with a smaller color separation agreed better
with the SNLS data, but never having a $\chi^2$ probability relative to the
SK16 model larger than $4 \times 10^{-5}$. 
Any SN~Ia model that has two narrow features in the $B$-peak $u-v$ color separated
by more the 0.25 mag does not agree well with the data.  
Adding additional color smearing to the Milne-like model would make
it agree better with the data, but would be inconsistent with the 
narrow widths of the color peaks seen in \citet{Milne2015}.

\section{Conclusions}
\label{sec:conclude}

Our analysis of the SDSS and SNLS supernova
surveys does not agree with the observations reported in
\citet{Milne2015}.  We do not observe two distinct red
and blue samples in the rest frame near ultraviolet brightness of
SN~Ia at $B$-peak.  Rather, we see a broad distribution
well described by a combination of SN~Ia color
variations and extinction as given by the SALT-II model with
a data derived distribution of its parameters, the SK16 model.
Our simulations of the two surveys show that we can distinguish
between the two $u-v$ color models, and our analysis of 158 light curves,
specifically the 99 from the SNLS, show no evidence for the
distinct $u-v$ peaks reported in \citet{Milne2015}.

\section*{Acknowledgments}

We thank Peter Milne and Ryan Foley for useful discussions about their work.
This work was supported in part by the Kavli Institute for Cosmological Physics at
the University of Chicago through grant NSF PHY-1125897 and an endowment from the 
Kavli Foundation and its founder Fred Kavli.
R.K and D.S, gratefully acknowledges support from NASA grant 14-WPS14-0048.  
D.S. is supported by NASA through Hubble Fellowship grant HST-HF2-51383.001 
awarded by the Space Telescope Science Institute, which is operated by the 
Association of Universities for Research in Astronomy, Inc., for NASA, under 
contract NAS 5-26555.




\bibliographystyle{mnras}




\section{Appendix}
\label{App}
%

The list of the SNe~Ia used in this analysis.  

The SNLS ID:
03D1au;
03D1aw;
03D1fc;
03D4au;
03D4cz;
03D4dh;
03D4dy;
03D4gf;
03D4gg;
04D1hd;
04D1hx;
04D1jg;
04D1kj;
04D1oh;
04D1pg;
04D1rh;
04D1sa;
04D2an;
04D2fp;
04D2fs;
04D2gb;
04D2gc;
04D2iu;
04D2mc;
04D2mh;
04D2mj;
04D3co;
04D3df;
04D3do;
04D3fk;
04D3kr;
04D3nh;
04D4an;
04D4bq;
04D4fx;
04D4gg;
04D4ib;
04D4ic;
04D4in;
04D4jr;
04D4ju;
05D1cc;
05D1ck;
05D1dn;
05D1dx;
05D1ee;
05D1hm;
05D1ix;
05D1ke;
05D1kl;
05D2ab;
05D2bv;
05D2cb;
05D2ci;
05D2ck;
05D2dt;
05D2dw;
05D2eb;
05D2hc;
05D2he;
05D2ie;
05D2le;
05D2mp;
05D3cf;
05D3ci;
05D3dd;
05D3gp;
05D3jq;
05D3jr;
05D3lb;
05D3lc;
05D3mh;
05D3mx;
05D4af;
05D4av;
05D4bf;
05D4bm;
05D4cw;
05D4dt;
05D4ef;
05D4ej;
05D4ek;
05D4ff;
05D4fo;
06D2bk;
06D2ca;
06D2cc;
06D2ck;
06D3cc;
06D3el;
06D3et;
06D2gb;
06D3df;
06D3ed;
06D3em;
06D4ba;
06D4bo;
06D4co; and
06D4cq.

The SDSS CID:
1166;
1688;
2533;
4241;
4679;
5183;
5391;
5737;
5844;
5966;
6100;
6137;
6649;
6699;
6924;
7143;
7475;
7779;
8598;
9045;
9207;
10550;
12883;
13136;
13830;
13934;
14397;
14456;
14735;
15170;
15213;
15217;
15383;
15456;
15704;
15756;
15776;
16000;
16093;
16211;
16232;
16421;
16779;
16789;
17528;
18091;
18617;
18782;
19029;
19033;
19632;
19818;
20106;
20142;
20184;
20186;
20245;
20432; and
21042.


\bsp	
\label{lastpage}
\end{document}